# Experimental evidences of quantum confined 2D indirect excitons in single barrier GaAs/AlAs/GaAs heterostructure using photocapacitance at room temperature

Amit Bhunia[1], Mohit Kumar Singh[1], Y. Galvão Gobato[2], Mohamed Henini[3,4] and Shouvik Datta[1] *

[1]*Department of Physics & Center for Energy Science, Indian Institute of Science Education and Research, Pune 411008, Maharashtra, India*

[2]*Departamento de Física, Universidade Federal de São Carlos, 13560-905, São Carlos, SP, Brazil*

[3]*School of Physics and Astronomy, University of Nottingham, Nottingham NG7 2RD, UK*

[4]*UNESCO-UNISA Africa Chair in Nanoscience & Nanotechnology Laboratories. College of Graduate Studies, University of South Africa (UNISA), Muckleneuk Ridge, PO Box 392, Pretoria, South Africa*

*Corresponding author's email: shouvik@iiserpune.ac.in

## Abstract

We investigated excitonic absorptions in GaAs/AlAs/GaAs single barrier heterostructure using both photocapacitance and photocurrent spectroscopies at room temperature. Photocapacitance spectra show well defined resonance peak of indirect excitons formed around the Γ-AlAs barrier. Unlike DC-photocurrent spectra, frequency dependent photocapacitance spectra interestingly red shift, sharpen up and then decrease with increasing tunneling at higher biases. Such dissimilarities clearly point out that different exciton dynamics govern these two spectral measurements. We also argue why such quantum confined dipoles of indirect excitons can have thermodynamically finite probabilities to survive even at room temperature. Finally, our observations demonstrate that photocapacitance technique, which was seldom used to detect excitons in the past, is useful for selective detection and experimental tuning of relatively small numbers ($\sim 10^{11}/cm^2$) of photo-generated indirect excitons having large effective dipole moments in this type of quasi-two dimensional heterostructures.

## I. INTRODUCTION

Two dimensional (2D) exciton based semiconductor heterostructure devices are interesting systems for exploring many body condensed matter physics and for next generation coherent light emitters. It is possible to detect excitons at room temperature using any steady state method when the generation rate is higher than the recombination rate inside a quantum structure[1,2]. In general, various optical techniques have been used to investigate the physics of excitons. However, being charge neutral dipoles, excitons cannot conduct direct current unless these are dissociated into electrons and holes by an electric field. Nevertheless, excitons with non-zero dipole moments are expected to respond in capacitance measurements. Previous studies[3,4] have shown that both photocurrent and capacitance-voltage (C-V) spectroscopy techniques can be used to detect excitonic features. Labud et.al[5] and Pal et.al[6] investigated many body interactions of different excitonic species as well as signatures of indirect excitons in quantum dot samples using standard C-V measurements in presence of light. Recently, we also reported[7] a thermodynamic model of dielectric signatures of excitonic presence in quantum well laser diodes at room temperature using capacitance measurements. Besides, long lived[8] indirect excitons are expected to play important roles in excitonic Bose-Einstein condensation[9,10] ultra-low threshold, coherent, polaritonic light sources[11,12] and excitonic superconductivity[13,14]. However, one must note that long radiative life times may also indicate dark excitons[15] which are not necessarily indirect excitons.

Here we report on observations of resonant excitonic signatures of photo-generated, 2D quantum confined indirect excitons in single barrier GaAs/AlAs/GaAs quantum heterostructure using photocapacitance spectroscopy under different DC-biases and AC modulation frequencies (see from Section III onwards). These findings are explained in Section IV and are supported with further experimental results in Section V. In Section V.C, we discuss why 2D quantum confined indirect excitons can have thermodynamically finite probabilities to survive even at room temperature. We also describe how these excitonic dipoles are detected using photocapacitance spectroscopy which has never been used for detection of indirect excitons. Earlier people had usually created and probed indirect excitons, where external electric fields[16] were used to pull apart direct excitons formed inside coupled quantum wells. Such procedures often broaden the excitonic spectra and thereby compromise the nature of these excitons. In contrast, in this study, applied electric fields are mainly utilized to push electrons and holes towards each other to form spatially separated, indirect excitons across the Γ-AlAs potential barrier. This procedure results in uncharacteristic sharpening of photocapacitance spectral features of indirect excitons with increasing bias.

## II.     SAMPLE AND EXPERIMENTAL METHODS

Our sample was grown by molecular beam epitaxy on a semi-insulating GaAs (311)A substrate. It was followed by a highly doped 1.5 µm thick p-GaAs ($4\times10^{18}$ cm$^{-3}$) buffer layer which also served as the bottom electrical contact for our measurements. Afterwards a 100nm thick p-GaAs ($1\times10^{17}$ cm$^{-3}$), a 8 nm undoped AlAs quantum barrier having 100 nm undoped GaAs spacer layer on both sides, 100 nm n-GaAs ($2\times10^{16}$ cm$^{-3}$) and a 0.5 µm of highly n-doped GaAs ($4\times10^{18}$ cm$^{-3}$) capping layer were successively grown to complete the structure. Circular gold mesas with ring diameter of 400 µm and area ~$5\times10^{-4}$ cm$^2$ as top metal contacts were made to facilitate optical access from above.

Agilent's E4980A LCR meter with 30 mV of rms voltage at a frequency (*f*) of 1 kHz (unless mentioned otherwise) was used for photocapacitance measurements. A simple equivalent circuit consisting of capacitance (C) and conductance (G) in parallel was used to extract photocapacitance using the LCR meter such that |C| is always greater than |0.1×G/ω| where ω=2π*f*. DC-Photocurrent spectra were measured using both LCR meter in the DC mode and a Keithley 2611 source meter. A 1000 Watts quartz-tungsten-halogen lamp and Acton Research's SP2555 monochromator having 0.5 m focal length (with Δλ ~ 3.2 nm) was used as light source. Spectral response of lamp plus monochromator changes slowly and monotonically within the wavelength ranges used in our measurements.

In Figs. 1(a) and 1(b), DC current and capacitance $\left(C = \frac{dQ}{dV}\right)$ at a frequency of 5 kHz are respectively plotted against applied bias under dark and also under 870±3 nm selective photoexcitation from the top n-type side of the heterostructure. This 870 nm wavelength was chosen to photo excite carriers above the band edge of GaAs. We clearly observe significant photo induced enhancement in both DC-current and capacitance values under finite applied biases. However, we refrain from any quantitative estimates of carrier density of a particular charge from photocapacitance versus bias characteristics, which are strictly valid only under depletion approximation for a one sided junction. It further complicates as there is no uniform equipotential surface under photoexcitation with a ring shaped top contact and also due to the presence of both types of carriers across the AlAs barrier. It is likely that a surface Schottky barrier is created between the top metal contact and the top GaAs layer. This, however, goes away when sizable photo absorption biases that surface barrier in the forward direction. Henceforth, in this paper, we will only focus on the variation of sharp resonant peak in photocapacitance and photocurrent spectra under non-zero biases. It is well known that large

DC-currents can destroy quantum effects due to dielectric screening and averaging effects. Therefore, such sub-micro Ampere DC-photocurrents are also crucial for observing experimental signatures of quantum confined indirect excitons which are only a small fraction of the total number of photo generated electrons and holes.

### III.     RESULTS OF PHOTOCAPACITANCE AND DC-PHOTOCURRENT SPECTRA

Figures 2(a) and 2(b) show photocapacitance spectra for a fixed frequency (1.0 kHz) under both reverse and forward biases, respectively. At first, photocapacitance peak gradually increases with increasing bias levels. In fact, we observe distinctly sharp resonant peak features under reverse bias which point to the presence of steady state photo generation of excitons. Likewise, Figs. 2(c) and 2(d) show the DC-photocurrent spectra at different reverse and forward biases respectively. Peak like excitonic transitions are also visible in these photocurrent spectra under reverse bias. Usually, any peak feature in optical absorption spectra of a semiconductor indicates resonant excitonic transitions. As such, both photocapacitance and photocurrent spectra are dependent on optical absorption spectra. Nevertheless, space charge modifications, charge carrier transport and electronic defects can significantly influence the respective magnitudes and shapes of these spectra. Similar pronounced peaks may also appear in linear optical absorption spectra due to strong[17] light-matter interaction inside a resonant optical cavity and also in case of Fermi-edge singularities. However, this single barrier heterostructure and the moderate levels of photon flux (~$10^{15}$/cm$^2$) used in our room temperature measurements for photo generation certainly do not fall into the above categories. It is clear from Fig. 2(a) that finite, reverse bias larger than -0.2 V is required to observe prominent spectral signature of excitonic transitions.

We roughly estimate the applied electric fields at each bias by dividing the bias magnitude with the thickness (~208 nm) of the intrinsic region of the heterostructure. In addition, we understand that applied electric field across single barrier heterostructure is not only influenced by immobile ionized charges of the GaAs/AlAs/GaAs heterojunction but also by accumulated charge carriers across the AlAs barrier. The high density of charge carrier accumulations across the AlAs barrier was earlier[18,19] argued to be the main cause for the inverted band structure under high forward bias. This will be discussed in details in the next Section IV.

It is well known that strong electric field dissociates direct excitons[20,21] and broadens excitonic transitions in semiconductor quantum wells. However, we see from Fig. 2(a) that photocapacitance signature of excitonic resonances become sharper as the reverse bias is increased from -0.2 to -2.0 V. Excitonic peak heights in photocapacitance initially increase with increasing reverse (Fig. 2(a)) and forward (Fig. 2(b)) biases up to -1.0 V and +0.4 V, respectively. For greater bias levels, the photocapacitance values of excitonic peaks reduce and also strongly red shift. As mentioned above, excitonic peaks are more prominent and sharper under reverse bias than forward bias. On the contrary, excitonic peaks of photocurrent spectra (Figs. 2(c) and 2(d)) are much less prominent, change sign under opposite biases and do not red shift with increasing biases. All these differences indicate that excitonic populations contributing to photocapacitance and photocurrent spectra have somewhat different physical origins. We will explain the cause of these differences in the next section.

## IV. DISCUSSIONS BASED ON GaAs/AlAs/GaAs BAND ALIGNMENT AND FORMATION OF 2D INDIRECT EXCITONS

An illustration of the sample under illumination is shown in Fig. 3(a). AlAs has an indirect band gap with conduction band minima in the X valley. As such Γ- X valley mixing of GaAs/AlAs/GaAs heterostructure was investigated earlier[18,19,22] at ~2K under high forward bias. In Figs. 3(b) and 3(c), we depict the schematic band diagrams of GaAs/AlAs/GaAs in both Γ-Γ-Γ and Γ-X-Γ configurations. For any AlAs barrier thicker than 3 nm, it is well known that Γ-GaAs/Γ-AlAs/Γ-GaAs heterostructure can have large AlAs potential barriers in both conduction and valence bands which suppress direct tunneling. In the following paragraph we will describe why our experimental observations can be largely explained using the presence of this Γ-AlAs potential barrier.

In our case, it is likely that photo generated electrons and holes in GaAs are driven by applied bias to accumulate on both sides of Γ-AlAs potential barrier. We also show electron and hole accumulations across Γ-AlAs under both reverse and forward biases, respectively, in Figs. 3(b) and 3(c). Due to such charge accumulations, 2D electron-gas (2DEG) and 2D hole-gas (2DHG) are formed across the Γ-AlAs barrier in respective triangular quantum wells (TQWs) in two spatially separated GaAs layers[23]. Proximity of these 2DEG and 2DHG allows the formation of indirect excitons by Coulomb attractions as shown in Fig. 3.

However, if we still want to consider indirect excitons within the X band of AlAs[18], which have a much smaller conduction band offset at Γ-GaAs/X-AlAs/Γ-GaAs heterostructure, then significant inter valley electron injection can also occur under applied bias. Such electrons can then recombine with holes of 2DHG. At temperatures ~2K, this was shown to have a dominant contribution to conductivity and electroluminescence. However, we wish to emphasize the following points in connection with our room temperature measurements. Firstly, any significant Γ-X-Γ inter valley electron tunneling, if present at room temperature, can deplete indirect excitons through radiative recombination and subsequently enhance tunneling conductivity even at smaller biases. Moreover, this Γ-X-Γ electron transport through inter valley tunneling is highly dependent[18] on applied bias and must show excitonic stark effects due to effective quantum confinement of electrons within X-GaAs/X-AlAs/X-GaAs band configuration. However, we do not observe any red shift of excitonic photocurrent spectra with applied bias (Figs. 2(c) and 2(d)). In Section V.C, we will confirm that the estimated diameter of indirect excitons is sufficiently large, which approaches the thickness of this Γ-AlAs barrier under higher biases. Likewise, it is still possible that Γ-X-Γ inter valley electron transport can dominate the tunneling current across AlAs barrier. However, thermal activation of charge carriers over the AlAs barrier may also contribute substantially to photocurrents measured at room temperature. This will be discussed later in Section V.D. Finally, it is also improbable that this Γ-X-Γ inter valley tunneling current can be a large fraction of the overall photocurrent as majority of photo generations take place in top GaAs layers. Therefore, we will restrict our subsequent discussions on indirect excitons formed only across this Γ-AlAs barrier. We will also discuss a qualitatively different origin of DC-photocurrent in the next paragraph.

It is also important to note that most of these photocurrents originate from excitonic photo absorptions in the top n-type GaAs layer (Fig. 2(c)). A sizable fraction of these excitons possibly form away from the GaAs/AlAs heterojunction. Furthermore, direct excitons in bulk GaAs are known to have a binding energy of ~5meV. Therefore, most bulk excitons generated within a depth of a few microns from the top n-type layer can either thermally dissociate or drift apart under the applied electric field into electrons and holes at room temperature. Here, these electrons can quickly flow out of the heterojunction under reverse bias. However, holes from these remote excitons cannot overcome the high $\Gamma$-AlAs barrier and hardly contribute to DC-photocurrent under lower biases. Therefore, unlike photocapacitance, bias driven flow of carriers out of the GaAs/AlAs/GaAs heterojunction contributes to DC-photocurrent. Nevertheless, in presence of weaker quantum confinement due to reduced band bending, the number of indirect excitons formed under forward bias is smaller than that of reverse bias. Thus, sharp excitonic absorptions in both photocapacitance and photocurrent spectra under forward bias are barely visible. Moreover, we expect photo generated holes in n-GaAs layer to move away from the heterojunction under forward bias. In addition, these holes with larger effective masses are also minority carriers in the top n-GaAs layer. Hence, these cannot flow out of the heterojunction efficiently like electrons in case of reverse bias. As a result, we see comparatively smaller photocurrent signal under forward bias. We, therefore, emphasize that measured photocapacitance is likely connected with indirect excitons formed within the 2DEG and 2DHG which are on opposite sides of the $\Gamma$-AlAs barrier. Whereas all other excitons photo generated elsewhere inside the heterostructure can contribute to DC-photocurrent without any visible red shift. We will now present additional experimental results and analyses in Section V to support all these ideas.

# V. ADDITIONAL EXPERIMENTAL RESULTS ON INDIRECT EXCITONS AND THEIR FREQUENCY RESPONSE

## A. Estimation of dipole moments of indirect excitons from bias dependence of photocapacitance spectra

Under applied bias voltages, excitons inside a quantum structure can show quantum confined Stark Effect (QCSE)[24]. We also observe a sizable red shift of excitonic photocapacitance peaks under reverse bias as shown in Fig. 4(a). Peak energy shifts by ~14 meV within the applied bias range. Recently[25] it was reported that red shift of the peak excitonic energy per applied voltage (i.e. $\delta E_{IX}/V$) remains ~10 meV/V. Similarly, we find $\delta E_{IX}/V$ ~8 meV/V. Inset of Fig. 4(a) displays spectral red shifts of excitonic photocapacitance peaks in the presence of electric field under forward bias. Variation of peak energy (E) of excitonic photocapacitance signal with applied electric field ($\vec{F}$) is fitted with the following relation[26].

$$E = E_0 + \vec{p}.\vec{F} + \beta \vec{F}^2 \quad (1)$$

where $E_0$ is zero field energy, $\vec{p}$ is permanent dipole moment of indirect excitons, and β is polarizability. Using Eq. (1), we obtain reasonably high values of dipole moments as $\vec{p} = (+7.5 \pm 0.7) \times 10^{-28}$ C.m and $\vec{p} = (-2.0 \pm 0.2) \times 10^{-28}$ C.m for reverse and forward biases, respectively. The signs in front of $\vec{p}$ take care of respective directions of excitonic dipoles. Interestingly, dipole moments ($\vec{p}$) and applied electric fields ($\vec{F}$) are always in the opposite direction. Such anti-parallel alignments of $\vec{p}$ and $\vec{F}$ are shown in Figs. 3(b) and 3(c). These are certainly not the lowest energy configurations for these excitonic dipoles. Therefore, this change in sign of effective built-in dipole moment with changes in the direction of bias fits well with our band diagrams (Figs. 3(b) & 3(c)) which display '*inverted*' dipoles of indirect excitons. We estimate polarizability as $+4.5 \times 10^{-35}\ C\ m^2 V^{-1}$ and $+1.2 \times 10^{-35}\ Cm^2 V^{-1}$ under reverse and forward biases, respectively. We can approximate Eq. (1) as,

$$E = E_0 + |\vec{p}|\left(\hat{p} + \frac{\beta}{|\vec{p}|}\vec{F}\right).\vec{F} = E_0 + \vec{p}'_{effective}.\vec{F} \quad (2)$$

here β is assumed to be independent of $\vec{F}$ and $\vec{p}$. In addition, $\hat{p}$ is the unit vector of respective dipole moment $\vec{p} = |\vec{p}|\hat{p}$ under both types of biases. From band diagrams in Figs. 3(b) and 3(c), one can notice that the term $\left(\frac{\beta}{|\vec{p}|}\vec{F}\right)$ is always in the opposite direction to $\vec{p}$ in each case. As a result, the effective dipole moment $\vec{p}'_{effective}$ decreases with increasing applied electric field under both reverse bias and forward bias. This can only happen if increasing applied electric field can bring electrons and holes of these dipoles of indirect excitons closer as per the above band diagrams. Therefore, the above analyses also corroborate our discussions in Section IV and provide an understanding on how photocapacitance is able to detect the indirect excitons by sensing their dipolar contributions. Such sizable red shift also points towards substantial quantum confinement of constituent electrons and holes in the 2DEG and 2DHG, respectively. We will elaborate later how such quantum confinement also raises the binding energy of these indirect excitons, which will be illustrated in the inset of Fig. 5(b).

### B. Effect of tunneling on the population density of indirect excitons under higher biases

Figure 4(b) shows the magnitudes of excitonic photocapacitance peaks under reverse biases. It increases up to a bias of -1.0 V and then decreases for higher biases. Similar behavior is observed for forward biases in inset of Fig. 4(b). This behavior is expected when the Γ-AlAs barrier is subjected to large biases at room temperature. It allows photo generated charge carriers to tunnel through AlAs. This tunneling can subsequently reduce the population of indirect excitons which is reflected in the variation of peak value of excitonic photocapacitance (Fig. 4(b)). To confirm this, we verify the presence of Fowler-Nordheim (FN) tunneling at higher biases. This can be established by checking whether the DC-photocurrent density obeys the following equation.

$$J_{FN} = \frac{q^3}{16\pi^2 \hbar \Phi_b} F^2 \exp\left(-\frac{4}{3}\frac{\sqrt{2m^*}}{\hbar q}\Phi_b^{\frac{3}{2}}\frac{1}{F}\right) \qquad (3)$$

where $J_{FN}$ is photo current density, q is the electron charge, $\hbar$ is reduced Planck's constant, $m^*$ is effective mass of electron (hole) in GaAs, $\Phi_b$ is the effective barrier height at the GaAs/AlAs interface and $F$ is applied electric field across heterostructure. Figure 4(c) shows the Fowler-Nordheim plot for both forward and reverse biases. Characteristic linear regions at higher biases indicate the presence of tunneling, which also coincide with bias induced reduction of excitonic photocapacitance peaks. Therefore, it establishes that the tunneling process reduces the density of indirect excitons resulting in a significant decrease of the excitonic contribution in photocapacitance spectra (see Figs. 2(a), 2(b) and 4(b)). Remarkably, tunneling through Γ-AlAs affects excitonic photocapacitance and photocurrent spectra in a different way. Electrons and holes of indirect excitons can preferentially tunnel at higher biases due to their proximity to the AlAs tunnel barrier. However, this tunneling current likely contributes a negligible fraction to the overall increase of bulk photocurrent signal and it does not shift the excitonic peak with increasing biases. It is worth noting that electrons having low effective mass ($m^*$) can easily penetrate the AlAs quantum barrier as compared to holes. This is due to the fact that the wave function inside AlAs tunnel barrier varies with position ($x$) as $\sim(\exp(-\kappa x), \kappa \sim \sqrt{m^*})$.

## C. Additional confirmatory evidences of indirect excitons from photocapacitance spectra

In Fig. 5(a), we plot the photocapacitance versus photon flux for -1.0 V reverse bias and 877 ± 3 nm photoexcitation wavelength. This selected wavelength corresponds to a maximum excitonic absorption under -1.0 V. Here we assume that photocapacitance signal is proportional to $\eta N^\alpha$ where N is the photon flux, $\eta$ is the fraction of incident photons which converts to indirect excitons and $\alpha$ is the power law exponent. Dipolar dispersion of complex permittivity under simplistic driven, damped, simple harmonic oscillator model is always proportional to the number density of polarizable dipoles. We, therefore, assume that the photocapacitance signal has one-to-one correspondence with areal density of these dipoles of indirect excitons ($n_{IX}$). In addition, we find that value of α is around one. This indeed supports our explanation that photocapacitance is directly proportional to the density of dipolar indirect excitons. The fitted value of the fraction $\eta$ of 6.6×10$^{-5}$ then provides us with a reasonable upper limit of $n_{IX} \cong 1.2 \times 10^{11}/cm^2$ for a photon flux of $1.85 \times 10^{15}/cm^2$ used to measure the spectra. Presence of excitons other than indirect excitons in photocurrent spectra is similarly verified with a power law exponent which approaches one[27] (this plot is not shown here).

We now provide further evidences that indirect excitons are formed by a small fraction of photo generated electrons and holes which gradually come closer towards each other under increasing bias. In order to check this, we tried to analyze the red shifts of excitonic resonance in photocapacitance spectra from a different perspective. Here we assume that the observed red shifts mostly come from bias induced changes of these 2D quantum confined indirect excitons. Subsequently, we estimate the change in average separation between these constituent electrons and holes of the 2DEG and 2DHG, respectively, using the following simplified formula,

$$E = E_g - \left(\frac{\hbar^2}{2\mu a_B^2(V)}\right) + \mathcal{O}(TQW) \qquad (4)$$

where E is the excitonic peak energy from the photocapacitance spectra, $E_g = 1.424\ eV$ is the bulk band gap of GaAs at room temperature, $a_B$ (V) is effective Bohr radius of indirect excitons which is a function of applied bias V, $\mu \sim 0.058 m_0$ is the reduced mass of excitons in GaAs, $m_0$ is the free electron mass. Here, we call the 2nd term within first bracket in Eq. (4) as the effective binding energy of these indirect excitons. The last term $\mathcal{O}(TQW)$ is, however, neglected in our estimation. This $\mathcal{O}(TQW)$ term may be dependent on separate quantum confinements of electrons and holes in two different TQWs and also on bias induced changes of these quantum wells. As a result, we note that our simplified analyses may underestimate this effective binding energy. However, for this study we will focus only on bias dependent changes of $a_B$ (V) as shown in Fig. 5(b). Here we observe that with increasing electric field under reverse bias, the effective excitonic diameter (= $2a_B$) of indirect excitons along the growth direction gradually decreases. In fact, we notice that the excitonic diameter approaches the average distance between the TQWs across the AlAs barrier. Binding energies of direct excitons are expected to reduce with increasing electric field if these are torn apart and subsequently ionize. However, in our case, qualitative estimates of effective binding energies of these indirect excitons increase with increasing reverse bias [inset of Fig. 5(b)]. This matches well with our interpretation of indirect excitons from the band diagrams shown in Figs. 3(b) and 3(c). Such characteristically opposite variation of effective binding energy with increasing applied bias further reinforces our claims that photocapacitance is selectively probing these indirect excitons whose size decrease under increasing reverse bias. On the other hand, from Figs. 2(c) and 2(d), we see that the effective Bohr radius of ~8.3 nm estimated from photocurrent peak remains relatively independent of the bias as excitonic photocurrent peak do not shift at all. It reaffirms that all these other excitons which contribute mostly to excitonic photocurrent are photo generated far away from the AlAs barrier. This is because remote excitons are not expected to significantly alter space charge regions near the barrier within the time scale of our photocapacitance response. Hence, these hardly contribute to steady state photocapacitance.

It is well known that quantum confinement increases binding energy of 2D excitons by nearly a factor of four[28,29]. Therefore, such relatively larger values of exciton binding energy as estimated from the data indicates that both electrons and holes of indirect excitons are possibly forced together by the external electric field to be confined in the TQWs as 2DEG and 2DHG, respectively. Moreover, dissociation of excitons into electrons and holes inside a solid is a many body statistical process with excitonic dissociation probability $\sim \exp\left(-\frac{E_b}{k_B T}\right)$, where $E_b$ is the exciton binding energy[30,31]. This clearly shows that even when binding energy $E_b$ is smaller than thermal fluctuation energy ($k_B T$), there are still finite, non-zero probabilities of having some bound states of excitons. This analysis also agrees well with our above observation that only a very small fraction (~$10^{-5}$) of photo-generated electrons and holes forms these indirect excitons. In addition, significantly smaller optical dielectric constant of AlAs as compared to GaAs also enhances their binding energy which in turn allows such quantum confined indirect excitons to survive even at room temperature. Therefore, it is remarkable that photocapacitance is sensitive enough to detect the presence of small areal densities (~$10^{11}$/cm$^2$) of indirect excitons which other processes such as photocurrent fail to observe. Accumulation of more electrons and holes in respective 2DEG and 2DHG can cause further band bending around these quantum wells under applied bias. Therefore, enhanced carrier accumulation can lead to the formation of 2DEG and 2DHG with stronger spatial confinements with increasing reverse biases. This additional spatial confinement under larger reverse biases also help these 2D indirect excitons to survive even at room temperature[1,2].

It is well known that fluctuations in quantum well width and inhomogeneity in alloy compositions can give rise to inhomogeneous broadenings of its optical spectra. Interestingly, inhomogeneous line widths of excitonic peaks of photocapacitance spectra shown in Fig. 5(c) decrease with increasing electric field associated with reverse bias. Strong presence of inhomogeneous broadening is exemplified in the inset of Fig. 5(c), where excitonic peak in photocapacitance spectra matches to a simple Gaussian line shape. Earlier photocurrent spectroscopy[32] was used to probe excitons within a quantum well, where electrons and holes are pushed in opposite direction in order to broaden the excitonic peaks with increasing electric fields. However, in our case, there is a potential barrier in the middle of GaAs/AlAs/GaAs heterojunction and resultant charge carrier dynamics are totally different. Here with increasing bias, photo generated electrons and holes of the indirect excitons come even closer towards each other as already observed in Fig. 5(b). Therefore, it results in unusual sharpening of excitonic photocapacitance with increasing reverse bias. This result again substantiates our interpretations of indirect excitons following the schematic band diagrams given in Figs. 3(b) and 3(c). In a way, it also validates the explanations given along with equation 4.

**D. Effect of AC modulation frequency on photocapacitance response**

Lately, we have reported[7] how a thermodynamic population of dipolar excitons in quantum confined laser diodes can respond to steady state differential capacitance measurements and shift towards higher frequencies with increasing bias values. Figures 6(a) and 6(b) show that excitonic photocapacitance values clearly increase and shift to higher frequencies with increasing bias under 870±3 nm selective photo excitation (increasing bias is shown by the arrows inside these figures). This supports the proposed[7] electrical signatures of formation of excitons. As stated above, tunneling reduces the number of accumulated electrons and holes in respective 2DEG and 2DHG regions, where these indirect excitons are located. As a result, photocapacitance signal also reduces once tunneling becomes significant at higher biases (inset of Figs. 6(a) and 6(b)). However, the range of frequency (~$10^2$ to ~$10^3$ Hz) for measured photocapacitance response is much smaller compared to frequency ranges (~$10^6$ Hz) reported in the earlier investigations[4,7,33]. These past studies had only probed direct excitons formed inside quantum structures. On the contrary, here electrons and holes of indirect excitons are physically separated by a large Γ-AlAs potential barrier which suppresses[34] free dipolar oscillation of excitonic charges. This is expected to cause considerable dipolar relaxation which can 'dampen' the oscillations of indirect excitonic dipoles during steady state photocapacitance measurements. In these experiments, dipolar relaxation kinetics of indirect excitons involves transitions between two states of opposite charge orientations separated by a large free energy barrier. Therefore, we use the following Arrhenius equation for thermally activated transitions over AlAs barrier

$$E_{Th} = k_B T ln(\nu/f) \tag{5}$$

where ν is the thermal pre-factor ~$10^{12}$ Hz, $k_B$ is the Boltzmann constant and T is the temperature in Kelvin. We calculate the effective activation barrier height ($E_{Th}$) to be ~0.56 eV at room temperature. We choose a frequency (*f*) of ~500 Hz where we observe a large photocapacitance response away from its final roll off at higher frequencies. This effective $E_{Th}$ nearly matches with the reported band offsets at Γ-GaAs/Γ-AlAs heterojunction where zero bias conduction and valence band offsets are $\Delta E_C$ ~0.8 eV and $\Delta E_V$ ~0.5 eV respectively. Figure 6(c) also shows that signature of excitonic resonance in the form of photocapacitance peak slowly vanishes above 1.0 kHz. Such low frequency response can only come from space charge modulation near the AlAs barrier, which directly affects the areal density of indirect excitons probed using photocapacitance. It is well known that space charge dipoles respond around this frequency range[34].

We also measured photocurrent vs bias under 870±3 nm selective photoexcitation. Using the photocurrent vs voltage plot, we obtain a reverse saturation value of this photocurrent density as $J_0 = 2\times 10^{-4}$ A/cm$^2$ by extrapolating the data below +0.5V. Furthermore, we use the following equation to approximately determine the barrier height ($V_B$)[35] for this photocurrent density.

$$V_B = q\varphi_B = k_B T \ ln\left(\frac{A^{**}T^2}{J_0}\right) \quad (6)$$

where $A^{**}$ is the reduced effective Richardson's constant which we approximate as ~120 Acm$^{-2}$T$^{-2}$, q is the electronic charge, $\varphi_B$ is the barrier potential and T = 300K. Effective barrier height estimated using this formula is 0.64 eV. We also estimate the built-in potential ($V_{Bi}$) from the doping profile of Γ-GaAs/Γ-AlAs/Γ-GaAs heterojunction using the following formula[35]

$$V_{Bi} = q\varphi_B = k_B T \ln\left(\frac{N_D N_A}{n_i^2}\right) \qquad (7)$$

where $N_D$ ($N_A$) is donor (acceptor) densities in n (p) type side of the heterostructure and $n_i \sim 10^{13} cm^{-3}$ is the carrier density in the undoped region. We find $V_{Bi} \sim 0.42$ eV.

Therefore, all these assessments of 'effective' barrier heights not only match with the above mentioned 'effective' activation energy barrier, but also comparable with expected band discontinuities at Γ-GaAs/Γ-AlAs/Γ-GaAs heterojunction. In a way, all these results fully support our above analyses on frequency dependent photocapacitance signatures of indirect excitons.

## VI. CONCLUSIONS

In summary, we have provided direct experimental evidences of optically generated, bias driven, inverted dipoles of 2D indirect excitons formed across 8 nm thin AlAs tunnel barrier in Γ bands of GaAs/AlAs/GaAs heterostructure. We have argued that quantum confinements of photo-generated electrons and holes in triangular 2DEG and 2DHG structures enhance both the excitonic binding energy and the thermodynamic probability of these indirect excitons to survive even at room temperature.

We have also analyzed characteristic differences in the electric field dependence of excitonic features in both photocapacitance and DC-photocurrent spectra. We have explained why valley selective electron and hole accumulation, mostly across Γ-AlAs band, can describe our experimental observations. In particular, we have shown that large, tunable dipole moment along the growth direction can be engineered to selectively probe, control and manipulate the indirect excitons by photocapacitance using bias and modulation frequency. Moreover, peak photocapacitance signal follows a reasonable upper limit of areal density of photo generated indirect excitons. Remarkably, the photocapacitance technique which was rarely used in the past to detect excitons, can be used to detect indirect excitons which constitute only a small fraction ($\sim 10^{-5}$) of the photo generated carriers.

Finally, this work opens up the possibility of experimental detection of indirect excitons using photocapacitance at room temperature. It also creates future possibilities for photocapacitance studies of excitonic-based many body condensed matter physics in heterostructures of 2D transition metal di-chalcogenides[13,14,36-38] using their well-known valley selective charge transport. Significant differences between excitonic signatures of photocapacitance and photocurrent may also be used to address either indirect or direct excitons. Therefore, we predict that all these above findings will also be helpful to design better excitonic devices[39].

## ACKNOWLEDGRMENTS

Authors acknowledge Department of Science and Technology, India (Research Grants # SR/S2/CMP-72/2012 and SR/NM/TP13/2016). We are grateful to Prof. B. M. Arora from IIT-Mumbai for his advices on band diagrams. AB and MKS are thankful to DST, India for Inspire Ph.D Fellowship and IISER-Pune Int. Ph.D Fellowship respectively. YGG acknowledges the financial support from the Brazilian agency Fundação de Amparo a Pesquisa do Estado de São Paulo (FAPESP) (Research Grant # 16/10668-7). MH acknowledges support from the UK Engineering and Physical Sciences Research Council.

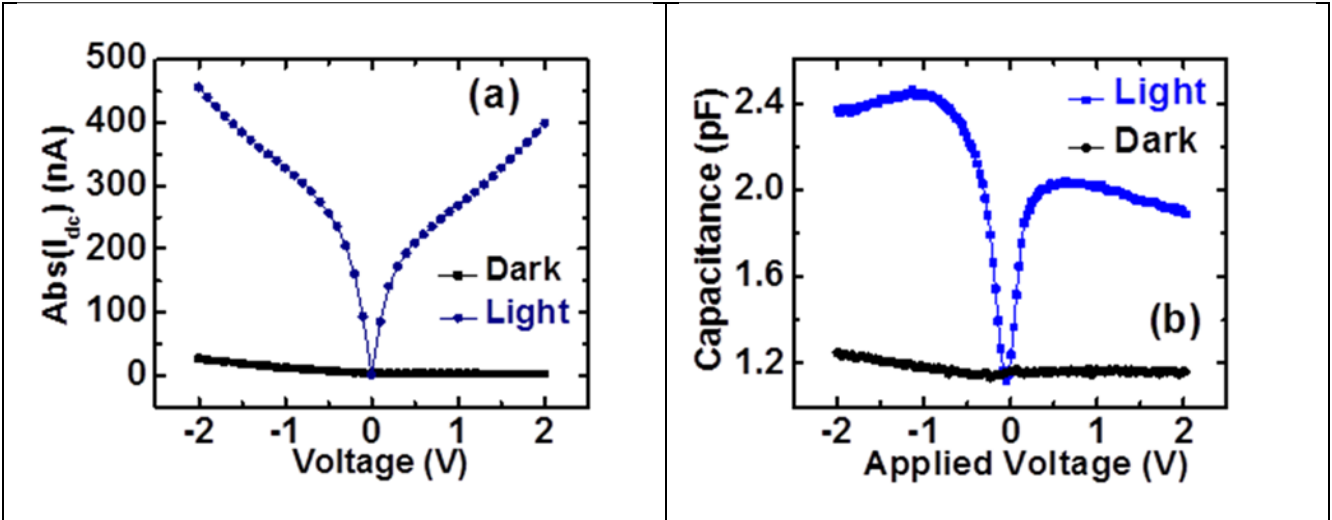

FIG. 1. (a) Absolute values of DC-current and (b) small signal capacitance at 5.0 kHz are plotted against applied bias under dark and under selective illumination of 870±3 nm, respectively. Significant photo induced changes are clearly visible in both cases. In this paper, we focus only on large photo induced changes in both capacitance and DC-current under non-zero applied biases for reasons mentioned in Section II.

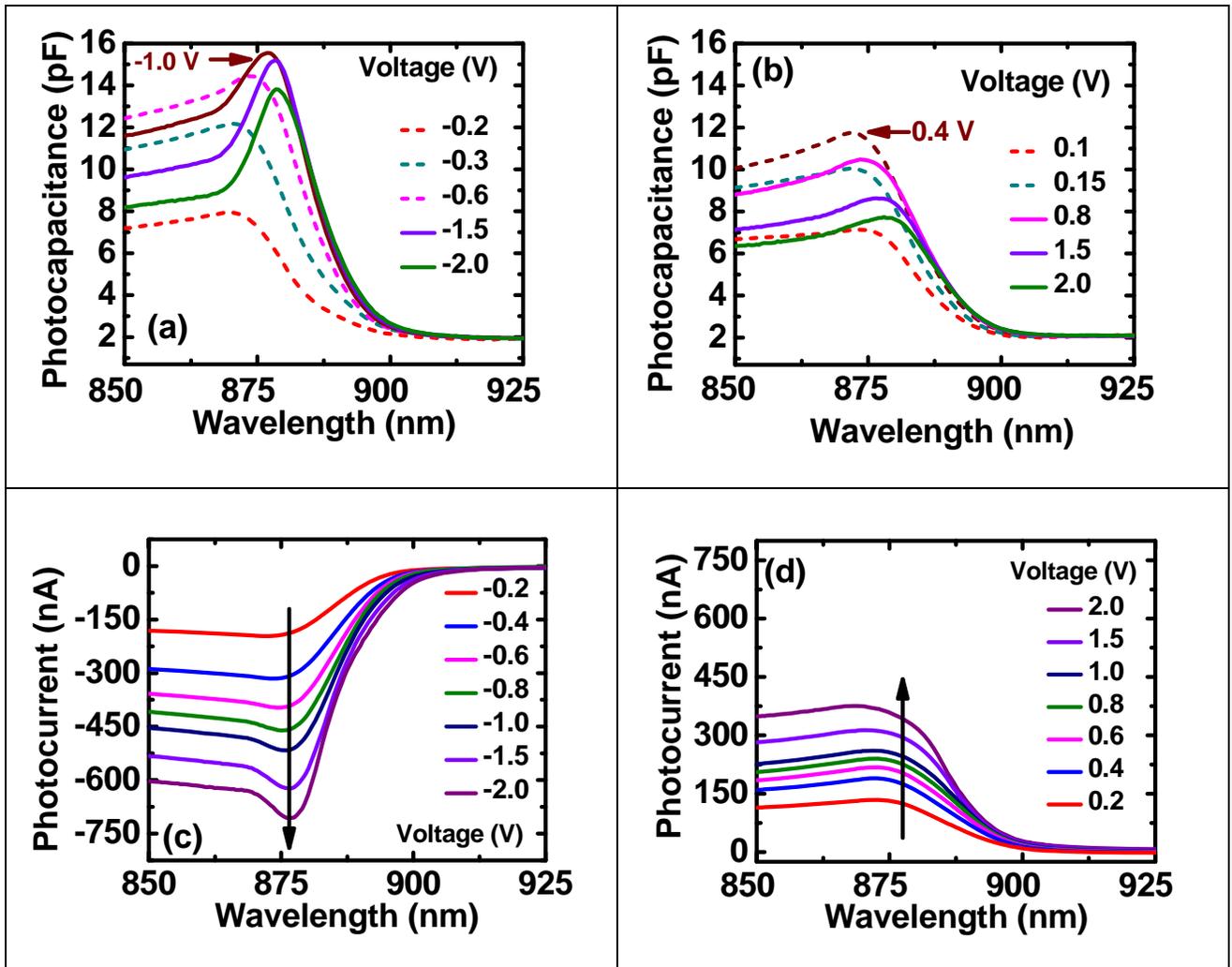

FIG. 2. Plots of bias dependent, 1.0 kHz photocapacitance spectra of GaAs/AlAs/GaAs single barrier heterostructure at room temperature under (a) reverse and (b) forward bias respectively. (c) and (d) Bias dependent photocurrent spectra plotted under reverse and forward bias, respectively. The arrows indicate the direction of increasing bias. Unlike photocapacitance spectra, photocurrent spectra hardly red shift and photocurrent signals always monotonically increase with increasing biases.

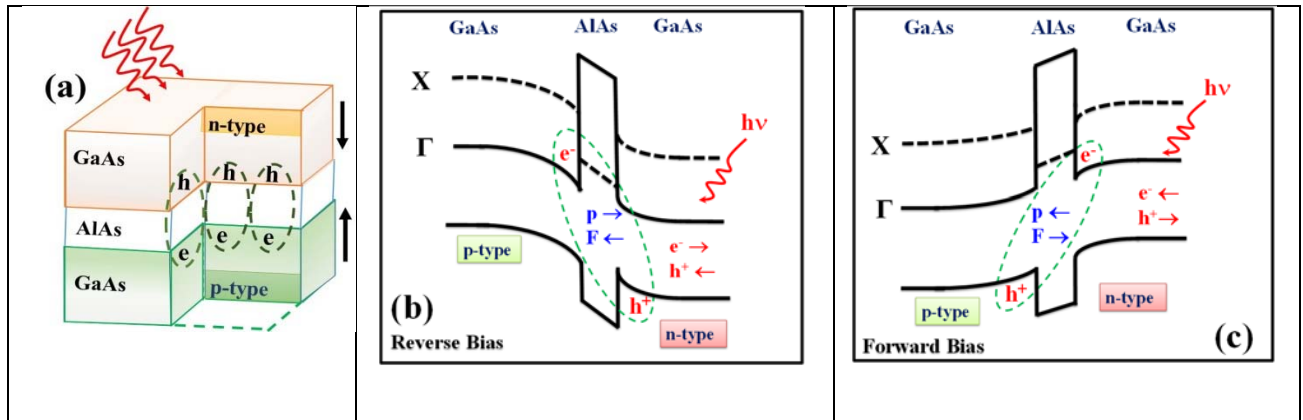

FIG. 3. (a) Depiction of photo-excited GaAs/AlAs/GaAs quantum heterostructure under reverse bias. Black arrows describe the bias driven motion of photo generated charge carriers which result in the formation of indirect excitons across the AlAs barrier. Schematic energy band diagrams of the heterojunction under reverse and forward biases are shown in (b) and (c), respectively and discussed in Section IV. Black dashed lines correspond to conduction band minima in X-band. Direction of motion for electrons ($e^-$) and holes ($h^+$) are indicated with red arrows. Direction of applied electric field ($\vec{F}$) and dipole moment ($\vec{p}$) of these indirect excitons are indicated with blue arrows. Green dashed lines are showing the outline of these indirect excitons in the above diagrams.

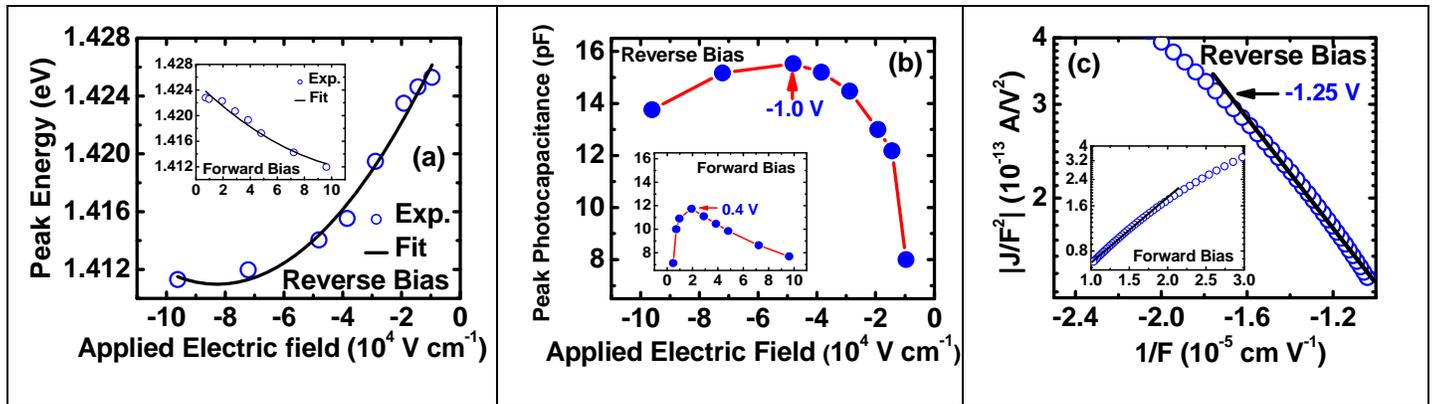

FIG. 4. (a) Excitonic peak energies estimated from photocapacitance spectra of Fig. 2(a) are plotted against applied electric field under reverse bias. The inset shows similar plot under forward bias. Black lines are fitted to extract the dipole moments and polarizability of indirect excitons. (b) Magnitude of excitonic photocapacitance peaks vary with the applied electric field under reverse bias and forward bias (Inset). Red lines are guide to the eyes only. (c) Straight line portions of Fowler-Nordheim plot at higher biases indicate the presence of tunneling. This happens around bias values beyond which excitonic photocapacitance peak magnitudes also decrease (Fig. 4(b)). Here the absolute magnitude $|J/F^2|$ is plotted along log-Y axis. Black straight lines are guide to eyes only.

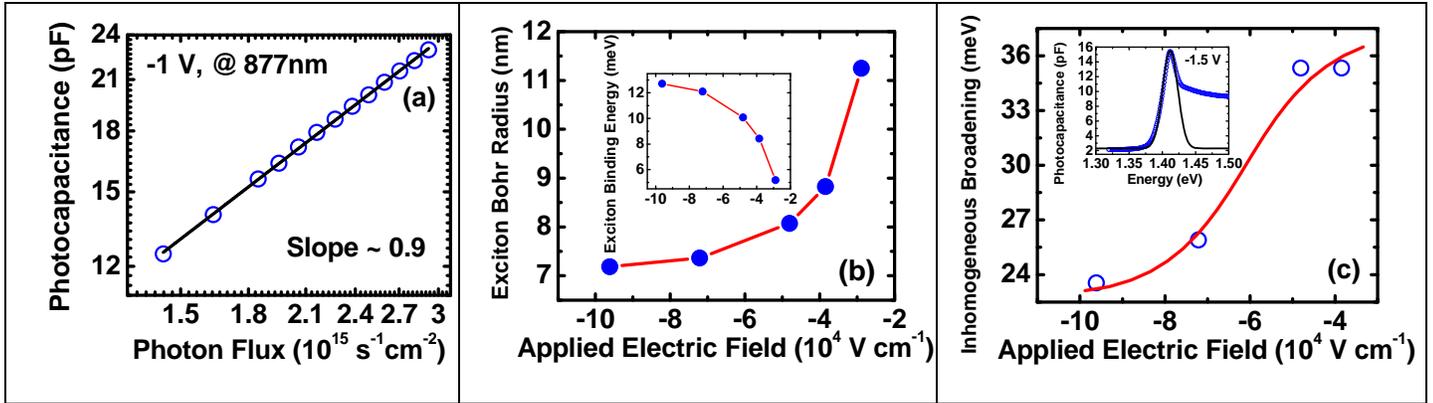

FIG. 5. (a) Power law fitting of excitonic photocapacitance peak vs photon flux in log-log scale. The slope as power law exponent (α) around unity indicates the presence of excitons and the direct connection of photocapacitance signal with the density of indirect excitons. (b) Estimated effective Bohr exciton radius ($a_B$) decreases with increasing electric field under reverse bias. The inset shows corresponding increase of effective exciton binding energy with increasing reverse bias. Solid red lines are guide to the eyes only. (c) Variation of inhomogeneous line widths of excitonic photocapacitance peaks decreases with increasing electric field under reverse bias. Red line is a guide to the eyes only. The inset shows that Gaussian line shape matches well with excitonic photocapacitance spectrum.

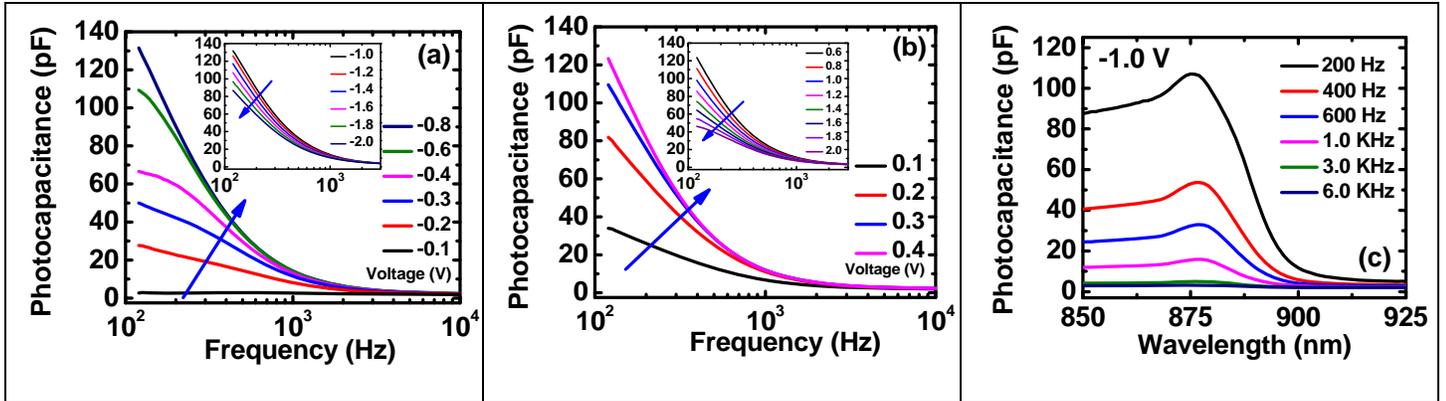

FIG. 6. (a) and (b) Photocapacitance as a function of modulation frequency at different reverse and forward biases, respectively. Photocapacitance response gradually increases and shifts towards higher frequencies with increasing biases until a threshold value is reached. Insets show further reduction of photocapacitance at higher biases once tunneling sets in. Blue arrows indicate the direction of increasing bias magnitudes. (c) Photocapacitance spectra for different modulation frequencies for a particular bias of -1.0 V.

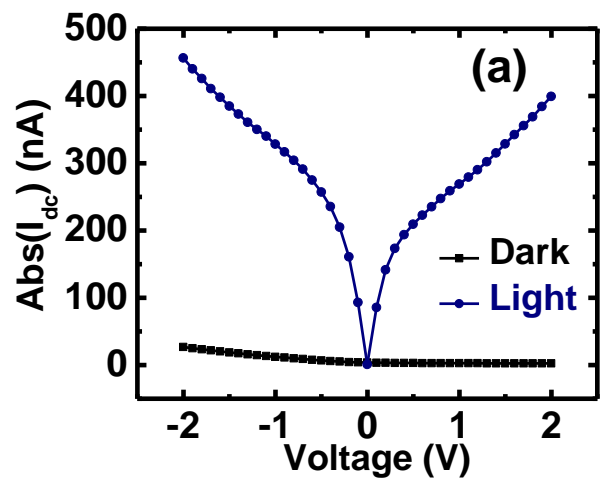 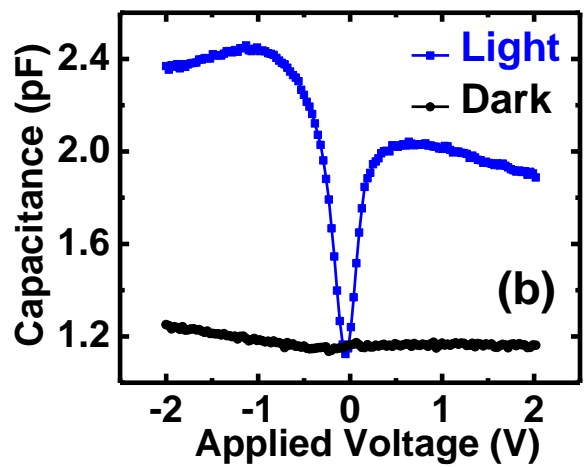

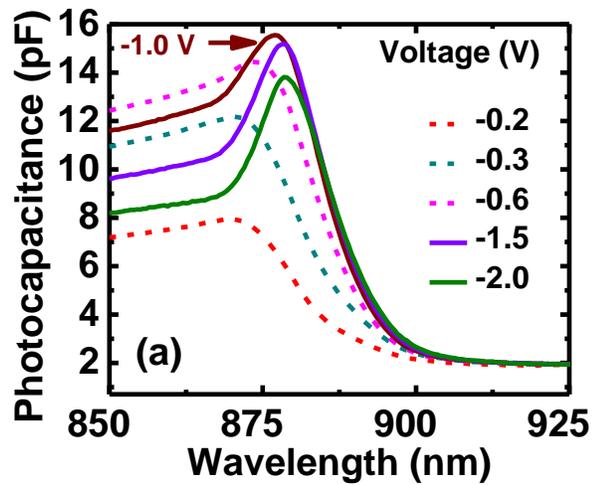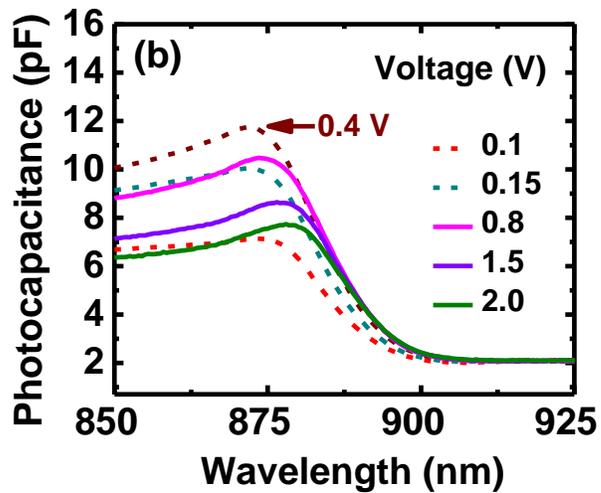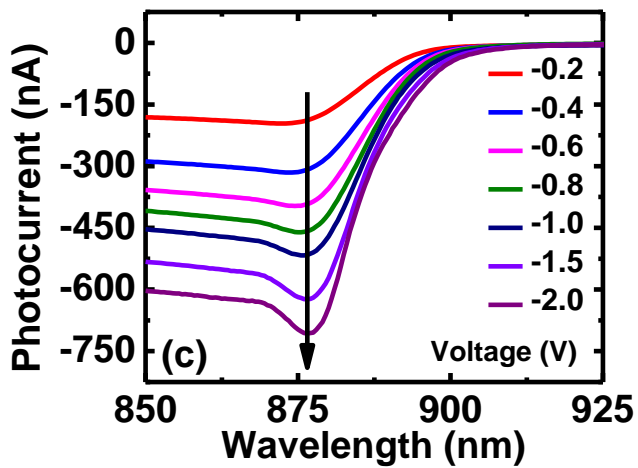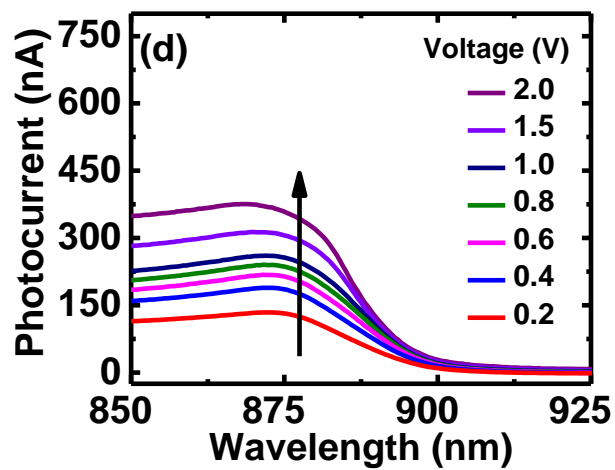

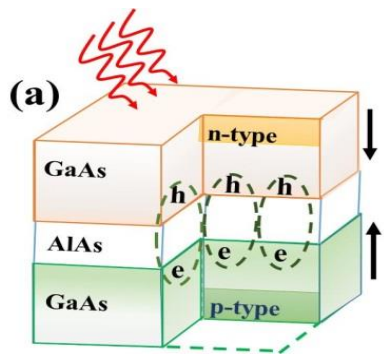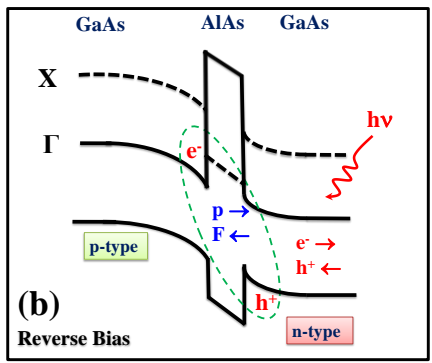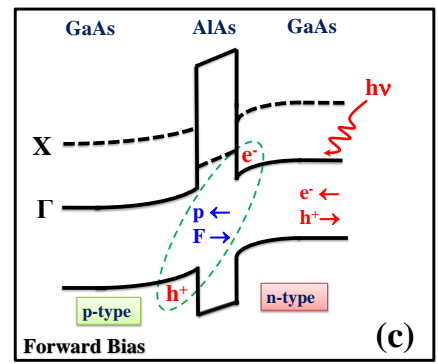

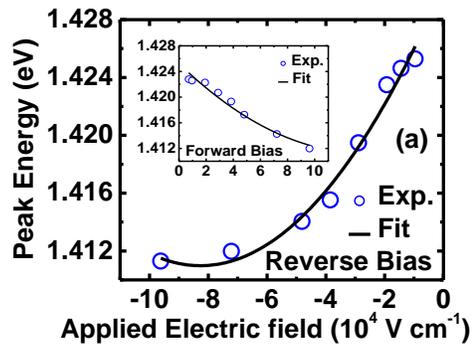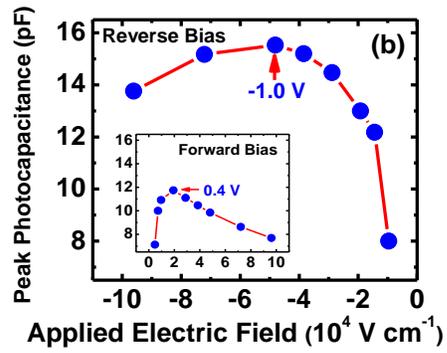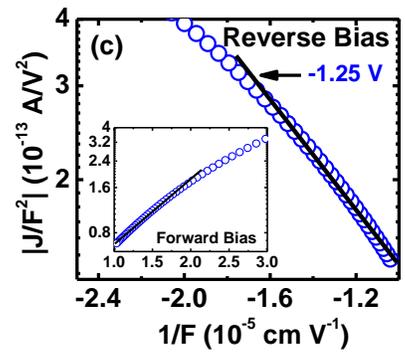

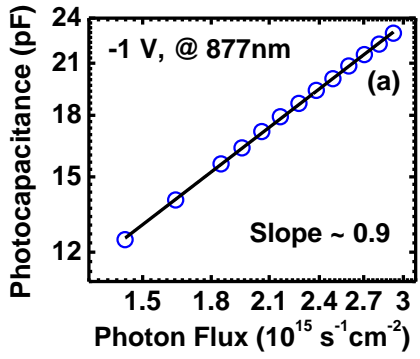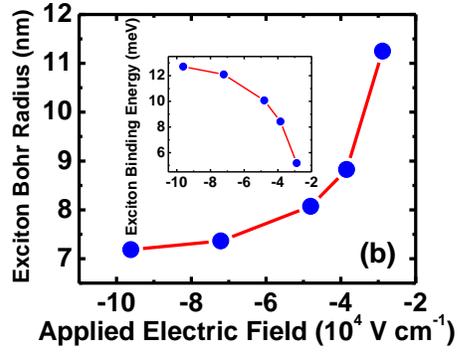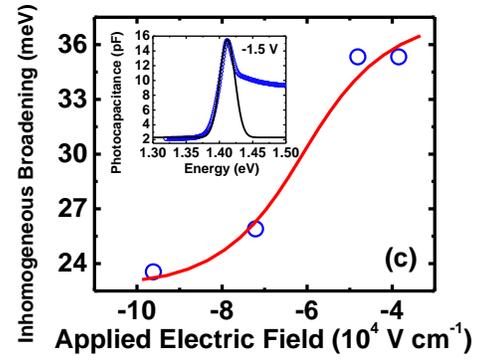

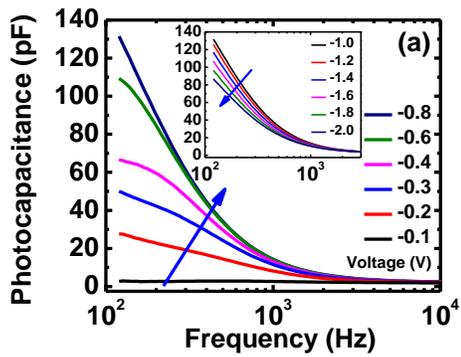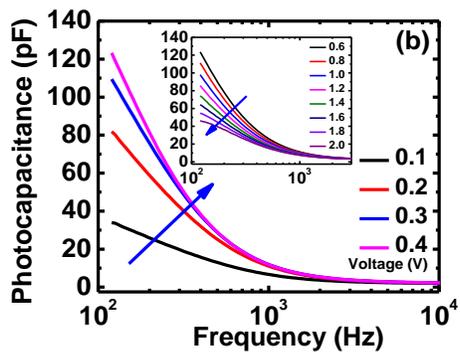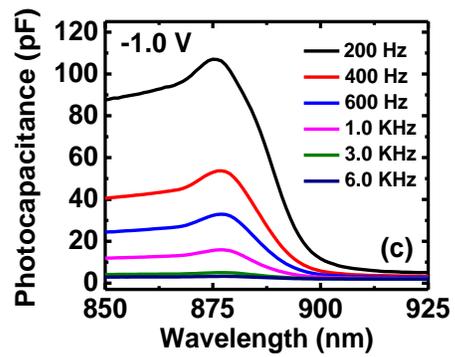